# An analysis of Internet Banking in Portugal:
## the antecedents of mobile banking adoption

João Pedro Couto
Business and Economics Department
University of Azores
Ponta Delgada, Portugal

Teresa Tiago
Business and Economics Department
University of Azores
Ponta Delgada, Portugal

Flávio Tiago
Business and Economics Department
University of Azores
Ponta Delgada, Portugal

*Abstract*—In recent years, mobile operations have gained wide popularity among mainstream users, and banks tend to follow this trend. But are bank customers ready to move forward? Mobile banking appears to be a natural extension of Internet banking, . Thus, to predict consumer decisions to adopt mobile banking, it's useful to understand the pattern of adoption of Internet banking (IB). This investigation seeks contribute to an expansion of the knowledge regarding this matter by researching Portuguese consumers' patterns and behaviors concerning the acceptance and use intention of IB as a foundation for establishing growth strategies of mobile banking. For data collection, we used an online "snowball" process. The statistical treatment used included a factor analysis in order to allow examination of the interrelationships between the original variables. The analysis was made possible by developing a set of factors that expresses the common traits between them. The results revealed that the majority of respondents did not identify problems with the use of and interaction with the IB service. The study generated some interesting findings. First, the data generally supports the conceptual framework presented. However, some points need to be made: (i) trust and convenience, from all the elements referenced in the literature as relevant from the client' perspective, continue to be a very important elements; (ii) the results did not support the paradigm that the characteristics of individuals affect their behavior as consumers, (iii) individual technological characteristics affect consumer adoption of IB service; (iv) consumer perceptions about the IB service affect their use, as reveal by the existence of three types of customers that show different practices and perceptions of IB; and (v) intention to use IB is dependent upon attitudes and subjective norms regarding the use of IB.[1]

## I. INTRODUCTION

Traditionally, banking was a simple branch-based operation and therefore the intensive street presence was considered a success critical factor. Over the past three decades, the technological evolution path impact the way services processes and the bank sector was no exception in the use of multiple technologies and applications used on daily base activities. The ICT adoption brings along a more and complete services offer, with high levels of security involve and cost reduction. For these reasons, more and more banks were investing in this form of customer interaction. Internet banking has created a paradigm shift, enriching banks offers and taking advantage of customers' predisposition to engage into a virtual relationship with their banks.

However, with the internet banking market increasing competition driven by consumer expectations and technology developments, IB left is central role and become a launching platform for the latest IT-driven bank-offer: mobile banking.

The literature review reveals that many studies were (and currently are being) conducted on the adoption of mobile banking by customers primarily on their perceptions about cost reduction, ease of use and convenience, reliability, and lastly but not the least, security and privacy. It also refers that these costumers had past experience with internet banking solutions. Therefore, this paper identifies the different segments of IB customers, paying special attention to trust and convenience as key concepts of consumer behavior, since they are common influences identified in IB and mobile-banking.

In this study, therefore, building upon prior research regarding consumer adoption and use of internet banking we try to identify the antecedents of mobile banking adoption, considering that the intention to use mobile banking is impacted by past IB experiences.

For practical reasons, this study focuses on the Portuguese consumers' patterns and behaviors' concerning the acceptance and use intention of internet banking. The data was gathered online and in order to test the hypothesis, a set of multivariate statistical analysis were performed.

The rest of this paper is organized as follows: Section 2 discusses the related literature reviewed for this research study; the subsequent sub-sections outline the conceptual model and experimental hypothesis on which the model is based; presents methodology and the discuss the empirical findings; Section 5 describes concludes the paper's results. The implications for industry as well as for research and limitations and scope for future research have been discussed in last Sections 6 and 7 respectively.

## II. RESEARCH BACKGROUND

It seems very cliché to start a work on internet defining the different paths of evolution of the home banking. However, when looking to this evolution and the adoption patterns follow by clients and its implications in banks strategies, it seems quite useful to quickly review this processes.

In the last two decades, the topic of home banking has gained its own space both in academics and business circles [1-

Acknowledgment: Funding for this work is granted by FCT – CEEAplA, Research Center for Applied Economics and data gathered by Ricardo Borges, to whom we thank for the sharing.

3]. However, home banking is a concept with a history greater than it appears to be, since it started with sales over phone process, followed by ATM and "dial-up" computers access and nowadays it's an umbrella combining of all of these with internet banking and mobile banking.

The explosion of internet usage and the huge funding in ICT initiatives, allowed the design of internet banking services offers, overcoming the spatial and time constrain of banking services, since it provides a 24/7 and global coverage [4].

The concept internet banking refers to the use of the internet as a remote delivery channel for banking services [5], allowing clients to access their bank and bank account and perform almost all the different types of transactions available through internet [4].

Since the nineties, the number of households with Internet access has increased dramatically, offering new markets possibilities for internet-based services such as internet banking [6]. In general, Europe has been and still is the leader in internet banking technology and usage. Particularly, in Portugal most banks offer internet banking solutions (See, BESnet was created in 1998 by BancoEspirito Santo).

There are two main reasons, listed in most research works, to traditional financial institutions engage internet banking activities. The first is related to lowering operational costs. The second regards, improving consumer banking services, increasing retaining consumers' rate and expanding consumer' share. Back in 1999, [4] suggested that IB segment was the most profitable business unit. More recent evidences confirm his conclusions and added that besides been more profitable this business model retained loyal and committed consumers when compared with traditional banking (ABA, 2004; Fox, 2005). For all these reasons, banks have recognized the importance to differentiate themselves from other financial institutions through new distribution channels. If in the last decade, banks were investing in internet banking.

Nowadays, mobile banking has been rapidly gaining popularity as a potential medium for electronic commerce. However, the diffusion of this channel is still in an early stage.

Both systems operate over the internet and enables customers 24 hour 7 day access to their account, and allows customers to conduct more complicated transactions, such as pay bills, applying for housing loan applications, online shopping, account consultation, and stock portfolio management.

The work of Dhungel, Acharya, &Upadhyay-Dhungel [7] remembers that banks invest in electronic channels to take advantage of its unique features as universal applicability, more speed to conduct transactions and less financial costs. But points to the existence of different stages of adoption of digital channels.

These authors also defend that when a new innovation becomes commercial feasible, the adoption of this new technology by potential users leads to its diffusion. They stressed on the importance of identifying a list of customers' profiles that will embrace new technologies in the earliest stages of its implementation.

Although, the development of the electronic banking supply, the number of mobile banking users is still very weak in comparison with the other e-banking services, such as internet banking and ATMS. Therefore, there is a need to understand the factors that influence intention to use mobile banking. Following a similar approach of the Dhungel, Acharya, &Upadhyay-Dhungel [7] work, it's relevant to determinate the key customer-specific factors that will predict actual digital behavior.

Few studies have focused on mobile value from the distinctive feature of a mobile technology perspective as an internet continuum technology, with specific customers' expectations [6]. Novel perspectives point to the replication of patterns from e-commerce to mobile solutions [8, 9].

In recent years, a multiplicity of theoretical perspectives have been applied to provide an understanding of the determinants of Internet banking adoption and use and more recently to understand mobile banking [1, 7, 10-13].

With the growing acceptance of internet banking, a constant analysis of customer behavior is needed, considering the factors affecting its adoption. In this field, often behavioral models are used, such as theory of reasoned action (TRA) [14] or the technology acceptance model (TAM) [15]. More recent studies employing a TAM-base theoretical lens have identified additional constructs that may be influential in internet banking service adoption: (i) online consumer behavior and online service adoption (channel knowledge, convenience, experience, perceived, accessibility and perceived utility; time savings; site waiting time; security, privacy and trust; cost; service quality); (ii) service switching cost (procedural, financial and relational); (iii) adoption of internet banking (convenience, service quality, perceived relative advantage, compatibility, trialability, complexity, demographics, gender, consumer attitudes and beliefs, security, privacy, trust, risk, needs already satisfied, familiarity, habit, convenience, adaptability, computer and technology confidence, knowledge and high levels of internet use at work) [14, 16-22].

With increasing technology adoption, bank performance is progressing steadily, therefore customer service is the area with major improvements made, which is a major advantage for users [19].

Regardless the importance of the subject, the investigation regarding European bank services is still scarce. When looking to the development of banking industry is clear that Portuguese banks are in the upstream of IT based solutions use. The questions that remain unanswered are how customers feel regarding internet banking and what influences their acceptance and adoption processes, including the adoption of mobile solutions.

### III. Framework and Hypothesis

Given the broad range of contributing theories and factors identified in the literature regarding banking in internet and mobile era, internet banking consumer behavior needs to study with a different approach to these factors. As Venkatesh et al. [21] recalled, the theory of reasoned action (TRA) from Fishbein and Ajzen [14] is still one of the most prominent theories regarding human behavior. According to this approach, behavior intention relies in two constructs: (i) the attitude toward behavior, which is "the positive or negative feelings about an individual's adoption of a particular behavior" (Fishbein&Ajzen, 1975, p. 216); (ii) and the subjective norm that is consider as "the perception of a person that most people who are important to him think he should or should not behave in a certain way "(Fishbein&Ajzen, 1975, p. 302). Since attitude resides in the mind, precedes and produces behavior and thus can be used to predict behavior (Yang and Yoo, 2004), it will reflect the personal characteristics of the individuals, such as gender, age or profession.

The principles of the theory of reasoned action were applied by Davis et al. [15] in the acceptance of IT by the individual, showing similar trends to other domains where TRA was applied. Driven from TRA, these authors present an instrument to predict the likelihood of a new technology being adopted within a group or individuals: technological acceptance model (TAM). Regardless the criticisms and the limitations found in TAMs, the model has been used in numerous studies seeking to gauge the determinants of behavior adoption and use of new technologies.

The positive results obtained from the use of this model in internet banking context (Daniel and Storey, 1998; Eriksson, Kerem, & Nilsson, 2005; Hernandez &Mazzon, 2007; Laukkanen, Sinkkonen, &Laukkanen, 2008; Mols, 1999; Prompattanapakdee, 2009; Qureshi & Khan, 2008; TeroPikkarainen, Pikkarainen, Karjaluoto, &Pahnila, 2004) and its non-application to Portuguese banking environment were the driving reasons for considering this model as the basis of their work. However, it was felt necessary to make some modifications, considering the inputs of the IBAM (internet banking adoption model) proposed by Alsajjan and Dennis [23] in order to enrich the model and a better match to the Portuguese reality. Therefore, special attention was given to trust and convenience as key concepts of online consumer behavior. For instance, from a psychology or relationship marketing perspective trust is the key element of relationships and in internet context has been found as important factor in the adoption decision process, especially in the internet banking context [3].

In light of the above discussion, the following hypotheses were constructed for testing in this study:

H1: Attitude towards the use of internet banking is dependent upon the behavioral beliefs relative to trust and convenience.

H2: Attitude towards the use internet banking is dependent upon individual demographic characteristics.

H3: Attitude towards the use internet banking is dependent upon individual technological characteristics.

H4: Intention to use internet banking is dependent upon attitudes towards the use of internet banking and subjective norms about the use of internet banking.

The hypotheses described reflect the study aims to explore the self-reported behaviors of online customers and their intention to use internet banking services. From the literature review that considers internet banking as the previous stage of mobile banking, it's expectable that the profiles found indicate future users of mobile banking [9].

IV. METHODOLOGY AND RESULTS

After the extensive literature survey, the research methodology has been centered on the already identified existing core variables. Therefore, and in order to validate the assumptions set, we applied a methodology consisting of four phases, namely, sample definition (1), questionnaire development (2), data collection (3), and statistical analysis (4). Based on the theorize model developed in the course of a detailed review of the related literature on user acceptance of technology and new technology diffusion, a questionnaire was compose as a measurement scale for the research, including two sections: (i) the first addresses issues of socio-demographic features of respondents, and (ii) the second adds a set of closed questions that capture differences in the perceptions of respondents to the use and adoption IB.

This study focuses on the Portuguese consumers for practical reasons and also because Portugal is one of the countries with more intensive electronic banking systems. Adding to it, the last two decades reflect a growing adoption of ICT in Portugal. According to the Eurostat, Portugal has an internet penetration rate larger than 61% and the Portuguese are the Europeans who most use the Internet to access bank services. However, the mobile banking penetration is still undersized, turning Portugal a good field for exploring the antecedents of mobile banking as a natural path after internet banking.

For data collection we used an online "snowball" process. The statistical treatment used, included a factor analysis in order to allow to study the interrelationships between the original variables, by developing a set of factors which expresses the common traits between them. Based on these factors we grouped the respondents according to their behavior relatively to the IB activities, carrying out a cluster analysis. To test the hypotheses we then applied tests of multiple means differences of to determine the existence of distinct patterns between the groups obtained in cluster analysis.

The final sample obtained comprises a set of 277, where 193 were males and 84 females, with 44% of the individuals between the ages of 25 to 34 years. The results revealed that the majority of respondents did not identify problems with the use and interaction with the IB service. Since the fear that consumers feel about the safety of online transaction is one of the biggest inhibitors of IB use [24], we tried to check the perception that the respondents had on the security provided by IB service of their bank, with the majority of respondents demonstrating confidence in the service using IB.

In order to test the hypothesis, a set of multivariate statistical analysis were performed. First, a factor analysis was applied, to reduce the number of variables on customer's perceptions of the service.

The results permitted the extraction of two factors, representing 63.41% of the total variance explained. The suitability of the technique is supported by the statistical significance of Bartlett's test and Kaiser-Meyer-Olkintest (KMO). Considering the variables associated with each

*(IJACSA) International Journal of Advanced Computer Science and Applications,*
*Vol. 4, No. 11, 2013*

component, the first factor was designated by "Trust" and reflects the way users perceive confidence and confidentiality and data security, the second factor, was designated by "Convenience", and is mainly associated with ability to perform all kind of operations, faster and at any given time.

TABLE I.     FACTOR ANALYSIS

| Rotated Component Matrix(a) | 1 | 2 |
|---|---|---|
| Offers na trustable service | 0,870 | 0,245 |
| Ensures clients privacy | 0,865 | 0,264 |
| Is a safe storage system | 0,833 | 0,312 |
| Has a positive reputation | 0,813 | 0,211 |
| Offers conveniente location | 0,155 | 0,744 |
| Offers conveniente hours | 0,395 | 0,709 |
| Offers the necessary operations | 0,368 | 0,707 |
| Offers more rapid transactions | 0,233 | 0,676 |
| Offers convenient information | 0,182 | 0,625 |
| Reduces the costs of transactions | 0,120 | 0,624 |

With the factors found, a cluster analysis was performed, using the K-means method. The solution showed three clusters, as can be seen in the following table. Given the factors found with higher incidence in these three clusters they were designated: Intermediate users (clusters 1), Full Users (cluster 2) and Basic users (cluster 3) (Fig. 1)

TABLE II.     CLUSTER ANALYSIS

| Dimension/ Cluster | Intermediate Users (n= 105) | Full Users (n= 130) | Basic Users (n= 42) |
|---|---|---|---|
| **Trust** | ,240 | ,323 | -1,599 |
| **Convenience** | -,961 | ,760 | ,050 |

Aiming to determine the existence of differences in how individuals perceive the activity of Internet Banking, based on their socio-demographic characteristics, a chi-square test was used crossing the cluster membership and these variables, namely sex, age, location, education background and profession. The results suggest the absence of significant differences between individuals with different demographic characteristics.

The existence of differences arising from the level of technological expertise of the individual was also analyzed, in line with that suggested by Eriksson et al. [25]. The results suggest the existence of significant differences in how respondents use the IB, depending on the experience level of Internet use, and IB use and the intensity of usage.

To understand how the three clusters obtained, are different from each other, in relation to other key issues, a variance analysis was performed as well as a set of tests of multiple comparisons of means. Regarding the attitude towards IB services we found that the respondents included in group 2 (full users), show a more favorable attitude on the various dimensions analyzed of attractiveness, sensibleness, beneficial and valuable service usage.

TABLE III.     ANALYSIS OF VARIANCE: ATTITUDE

| | Sum of squares | df | Mean Square | F | Sig. | Means Difference |
|---|---|---|---|---|---|---|
| Attractive to customers | 35,333 | 2 | 17,667 | 39,602 | 0 | 2>1,3 |
| A sensible option | 38,15 | 2 | 19,075 | 40,541 | 0 | 2>1,3 |
| Beneficial to customers | 37,208 | 2 | 18,604 | 41,96 | 0 | 2>1,3 |
| A good idea | 34,748 | 2 | 17,374 | 46,966 | 0 | 2>1,3 |
| An option with added value | 26,636 | 2 | 13,318 | 12,895 | 0 | 2>1,3 |

We also asked respondents about the advantages, type of transactions performed online and the process of utilization. The results showed significant differences among the three groups, and the full users cluster showed a higher rating of this items when compared to other two groups.

V.     DISCUSSION AND CONCLUSIONS

With recent advances in information and communication technologies, internet-base commerce is having an increasingly profound impact on our daily lives, offering appealing and advantageous new services. As Cheng, Lam and Yeung [26] suggested banks are taking advantage of these opportunities and Internet banking is widely seen as the key and most popular delivery channel for banking services in the cyber age. The growing importance of IB is highly associated to cost reduction, but especially to the benefits in customer relationship.

From the customers viewpoint, the decision to use the IB may be motivated by convenience and ease of use, but to these may be added financial gains, since many institutions practice lower prices for their IB services, as well as some of the features inherent in the use of the Internet itself: easily and quickly access the information, available 24/7, and ubiquity. This suggests that there are notable trends in the use of mobile banking as an extension of internet banking, since the systems is supported by internet and the benefits are quite similar [9].

However, despite the wide range of attributes, not all customers of banking institutions adhere to these type of service and those who do not always employ the same reasons or use it with the same intensity.

Therefore, the aim of the work is to gain awareness of the various reasons explaining why Portuguese consumers are or not becoming internet banking users and to determine their profiles of use, since the most intensive users can be potential mobile clients. Based on a random sample of internet users, demographic, attitudinal, and behavioral characteristics of Internet banking (IB) users and non-users were examined.

The results confirm the first hypothesis which states that perceived usefulness and ease of use influence the intention and actual use of IB solutions.

The cluster analysis carried out revealed the existence of three types of clients that demonstrate distinct practices and perceptions of IB: basic users (1), intermediate users (2) and full users (3).

Basic users are characterized by a lack of confidence in the IB service to ensure their safety, while intermediate users, although do not appreciating the ease of use IB, are individuals who feel that the service is safe, and full users I. Users highlight the full indulgence that IB provides them, also have confidence in the security presented in IB, while the basic users

Based on the results obtained in the multivariate analyzes, we can infer significant differences in the way individuals perceive the use of IB and vary your level of experience of IB. We can observe that there are marked differences between the three clusters, and full users stand out for having an positive attitude to IB and intended use superior.

The second hypothesis was based on the concept that the characteristics of individuals affect their behavior as consumers, namely their demographic characteristics.

The results contradict this hypothesis and conclusions presented by SadiqSohail and Shanmugham [27]. Thus, we could not find empirical substantiation, in our sample, to validate the influence of demographic characteristics of individuals, such as gender, age, education, in the adopting the IB services.

The third hypothesis had as a reference the dimensions that have been extensively studied with regard to IB consumer behavior, and from which there is no reference for the Portuguese context, that is the technology acceptance by consumers. This measure the technological characteristics of consumers effect on the adoption of Internet Banking services.

In this case the data support the hypothesis, and the results show that the way individuals relate to the technological conditions its performance in terms of Internet Banking.

## VI. Contributions and Implications

The results generated some interesting findings. First, data support in general the conceptual framework presented. However, some mentions need to be made: (i) trust and convenience, from all the elements referenced in the literature as relevant from the client' perspective, continue to be a very important elements [2, 3, 28-30]; (ii) the results did not support the paradigm that the characteristics of individuals affect their behavior as consumers, which contradicts the conclusions of Sadiq and Shanmugham [27]; (iii) individual technological characteristics affect consumer adoption of internet banking service; (iv) consumer perceptions over the Internet Banking service affect their use, as reveal by the existence of three types of customers that show different practices and perceptions of IB; and (v) intention to use internet banking is dependent upon attitudes towards the use of internet banking and subjective norms about the use of internet banking. These last results are consistent with the literature, in particular the models of adoption of new technologies proposed by Eriksson and Such (2008) and by Alsajjan and Dennis [23]. These last findings also suggest that the intention to engage in a higher-technology relationship with the bank is dependent on past experiences in the digital context.

The main theoretical contributions of this study highlight is the importance of the evidence that the components associated with the TAM model - the perception of usefulness and ease of use are relevant in IB adoption.

For the sample analyzed, it was possible to demonstrate that perceived usefulness and ease of use is a determinant condition in the intention and actual use of IB solutions. It was also possible to verify the existence of the influence on the intensity of use of the IB solutions.

This research has, however, not supported the conclusions of past studies about the influence of the personal characteristics of individuals in the adopting the IB. We could not find validation for the concept that the intrinsic characteristics of consumers, such as age, sex, educational level and income affect the intensity and how they adhere to Internet Banking.

Once the demographic characteristics, contrary to what happened in most other studies, emerge as not influencing the compliance behavior, it seems urgent to deepen the knowledge of IB customers in order to make possible the development of services and communication strategies for more customized Internet banking services.

It was also demonstrated, for the sample analyzed, that of all the elements referenced in the literature as relevant in the perspective of the client, the trust remains a highly valued element as well as ease of use. Therefore the banks need to seek solutions that are relevant in a user's perspective to minimizing uncertainty and maximizing the use.

The bank branches and their own interactions with clients continue to be a key element, but banks should not neglect the service provided to customers in the digital environments.

Considering the aspects mentioned above, we can conclude that this work may contribute to the identification of new target segments and that the results obtained may be taken into account in the development of future communication campaigns and digital offers.

Given the nature of its operations and services, the banking sector emerged as an area prone to dissemination and adoption of technological innovations. Since the nineties that has witnessed the proliferation of home banking activities and more recently the Mobile and Internet Banking.

The use of the Internet as a channel of distribution and communication can be considered another step in the assimilation of technologies that began with the appearance of ATM machines. But if for banks using the IB has advantages for clients can also be a wide range of benefits.

From the point of view of the customer, the decision to use IB may be driven by convenience and ease of use. To these may be added the financial gains , since many institutions practice tables lower prices for their services IB , as well as some of the features inherent in the use of the Internet itself : easily and quickly access the information , available 24/7 and, ubiquity .

It appears, however, that despite the wide range of attributes, not all customers of banks adhere to this type of service and that they do not always employ the same reasons or with the same intensity.

## VII. LIMITATIONS AND FUTURE RESEARCH

Some useful preliminary insights are produced, however, leaving a considerable number of issues for future research, providing scholars with an opportunity to conduct further research in this field and practitioners with an opportunity to enhance adoption rates based on consumer behavior knowledge.A limitation of this study is the sample size, while it has sought to achieve a larger sample, financial and time constraints of a research prevented the full achievement of initial objective. Considering this situation, the sample was limited to a lower set of participants that somehow share a similar educational and cultural background. Future work should seek to extend the sample, as well as minimizing the effects of educational and cultural proximity.

It would also be interesting to assess the existence of attitudes and patterns of behavior, in the face of technologies, at the different regions of the country, which would allow the evaluation of the impact of the development of technological infrastructure on a regional basis and the effect on in the local customer's behavior.

Furthermore, the scope of the study could be broadened to include the comparison of buying behavior in physical versus virtual banking services, or even consider the inclusion of mobile banking.

From the standpoint of credit institutions, it also appears as an important aspects the assessment of the effect of substitution of channels, interactive communication and even loyalty and overall satisfaction of stakeholders. A line of future research could include the assessment of the impact of this type of service in the overall performance of banks.

Additionally, despite the Portuguese consumer's exhibit preferences similar to those observed in other developed countries, there are still elements that require a deeper future analysis.

APPENDIX

Table a1: Users Age

| Users Age | Frequency | Percent | Valid Percent | Cumulative Percent |
|---|---|---|---|---|
| 18 a 24 | 68 | 24,5 | 24,5 | 24,5 |
| 25 a 34 | 123 | 44,4 | 44,4 | 69,0 |
| 35 a 44 | 55 | 19,9 | 19,9 | 88,8 |
| 45 a 54 | 21 | 7,6 | 7,6 | 96,4 |
| 55 a 64 | 10 | 3,6 | 3,6 | 100 |
| Total | 277 | 100 | 100 | |

Table a2: Users Education

| Users Education | Frequency | Percent | Valid Percent | Cumulative Percent |
|---|---|---|---|---|
| Basic | 2 | 0,72 | 0,7 | 0,7 |
| College | 60 | 21,66 | 21,7 | 22,5 |
| Comunity College | 26 | 9,39 | 9,4 | 31,9 |
| University | 137 | 49,46 | 49,6 | 81,5 |
| Master | 48 | 17,33 | 17,4 | 98,9 |
| PhD | 3 | 1,08 | 1,1 | 100 |
| Total | 276 | 99,64 | 100 | |
| System | 1 | 0,36 | | |

Table a3: Users by Working Activity

| Users by Working Activity | Frequency | Percent | Valid Percent | Cumulative Percent |
|---|---|---|---|---|
| Self Emplyoed | 17 | 6,1 | 6,2 | 6,2 |
| Company Employed | 187 | 67,5 | 67,8 | 73,9 |
| Unemployed | 14 | 5,1 | 5,1 | 81,5 |
| Retired | 7 | 2,5 | 2,5 | 76,4 |
| Student | 45 | 16,2 | 16,3 | 97,8 |
| Other | 6 | 2,2 | 2,2 | 100 |
| Total | 276 | 99,6 | 100 | |
| System | 1 | 0,4 | | |

Table a4: Time of use of IB

| Time of use of IB | Frequency | Percent | Valid Percent | Cumulative Percent |
|---|---|---|---|---|
| Doesn´t use | 40 | 14,4 | 14,4 | 14,4 |
| less then 6 meses | 20 | 7,2 | 7,2 | 21,7 |
| From 6 moths - 2 years | 49 | 17,7 | 17,7 | 39,4 |
| More then 2 -years | 168 | 60,6 | 60,6 | 100 |
| Total | 277 | 100 | 100 | |

Table a5: Percentage of operatin using IB

| Percentage of operatin using IB | Frequency | Percent | Valid Percent | Cumulative Percent |
|---|---|---|---|---|
| Less then 15% | 69 | 24,9 | 24,9 | 24,9 |
| 15% - 29% | 23 | 8,3 | 8,3 | 33,2 |
| 30% - 44% | 23 | 8,3 | 8,3 | 41,5 |
| 45% - 55% | 18 | 6,5 | 6,5 | 48,0 |
| 55% - 69% | 23 | 8,3 | 8,3 | 56,3 |
| 70% - 94% | 58 | 20,9 | 20,9 | 77,3 |
| More then 95% | 63 | 22,7 | 22,7 | 100 |
| Total | 277 | 100 | 100 | |